\documentclass[conference]{IEEEtran}
\IEEEoverridecommandlockouts  
\usepackage[utf8]{inputenc}
\usepackage[T1]{fontenc}
\usepackage{url}
\usepackage{graphicx} 
\usepackage{amsmath}
\usepackage{cite}
\usepackage{svg}
\graphicspath{{./figs/}}
\DeclareGraphicsExtensions{.pdf,.jpeg,.png,.eps,.pdf}
\usepackage{multirow}
\usepackage{makecell}
\usepackage{threeparttable}
\usepackage{subfig}
\begin{document}
\title{Focused State Recognition Using EEG \\with Eye Movement-Assisted Annotation }
%
%
%

\author{
Tian-Hua Li$^1$, Tian-Fang Ma$^1$, Dan Peng$^2$, Wei-Long Zheng$^{1,*}$ and Bao-Liang Lu$^{1,2,*}$ \textit{Fellow, IEEE}
\thanks{This work was supported in part by grants from STI 2030-Major Projects+2022ZD0208500, National Natural Science Foundation of China (Grant No. 62376158), Shanghai Municipal Science and Technology Major Project (Grant No. 2021SHZD ZX), Shanghai Pujiang Program (Grant No. 22PJ1408600), Medical-Engineering Interdisciplinary Research Foundation of Shanghai Jiao Tong University "Jiao Tong Star" Program (YG2023ZD25, YG2024ZD25, YG2024QNA03), and GuangCi Professorship Program of RuiJin Hospital Shanghai Jiao Tong University School of Medicine.}
\thanks{$^1$Department of Computer Science and Engineering, Shanghai Jiao Tong
University. $^2$RuiJin-Mihoyo Laboratory, Clinical Neuroscience Center, RuiJin Hospital, Shanghai Jiao Tong University School of Medicine. }
\thanks{*Corresponding authors}
}

\maketitle

\begin{abstract}

With the rapid advancement in machine learning, the recognition and analysis of brain activity based on EEG and eye movement signals have attained a high level of sophistication. Utilizing deep learning models for learning EEG and eye movement features proves effective in classifying brain activities. A focused state indicates intense concentration on a task or thought. Distinguishing focused and unfocused states can be achieved through eye movement behaviors, reflecting variations in brain activities. By calculating binocular focusing point disparity in eye movement signals and integrating relevant EEG features, we propose an annotation method for focused states. The resulting comprehensive dataset, derived from raw data processed through a bio-acquisition device, includes both EEG features and focused labels annotated by eye movements. Extensive training and testing on several deep learning models, particularly the Transformer, yielded a 90.16\% accuracy on the subject-dependent experiments. The validity of this approach was demonstrated, with cross-subject experiments, key frequency band and brain region analyses confirming its generalizability and providing physiological explanations.
\end{abstract}

\section{INTRODUCTION}
\label{sec:Introduction}
With the evolution of computer technology, the synergistic advancement of bio-acquisition devices and deep learning models has significantly facilitated the development and widespread adoption of various models for mental state analysis. Recently, various recognition models for human mental states, particularly emotions, have been proposed, leveraging signals from eye movements and EEG. Lin \textit{et al.} \cite{0} validated the coupling correlation between eye movement and EEG signals. In the context of disease screening, Phothisonothai \textit{et al}. \cite{1} employed EEG and eye-tracking methods to identify the early stages of dementia. For emotion recognition, Zheng \textit{et al}. \cite{2} explored the discriminative capabilities of EEG and eye movement signals in distinguishing between five emotions. Subsequently, Li \textit{et al}. \cite{3} and Liu \textit{et al}.\cite{4} delved into the differentiation ability of EEG and eye movement signals for a broader set of five emotions.

The focused state is a mental state characterized by the concentrated attention of an individual on a task, activity, or thought. Recognizing and analyzing this state can be instrumental in applications like safety and mental illness screening. For instance, Ju \textit{et al.} \cite{5} utilized eye-movement signals of the focused state to screen for depression. Distinguishing between focused and unfocused states involves adhering to principles such as maintaining a gaze on a specific target and manifesting heightened brain activity. Contemporary research predominantly focuses on monitoring physiological states, exemplified by EEG signals, or employs behavioral observations, including eye tracking and movement gesture analysis. Cheng \textit{et al.} \cite{6} integrated eye movement signals and EEG signals to achieve improved classification performance for motor imagery tasks, while Casson \textit{et al.} \cite{7} annotated the moments of interest by using binocular gaze points and located the corresponding EEG signals for analysis. 

In the context of a focused state, wherein both eyes concentrate on a singular point, as opposed to an unfocused state characterized by binocular disparity resulting from mental distraction, we introduce an annotation method. Specifically, binocular disparity derived from eye movement signals can function as a reliable criterion for discerning a focused state. Furthermore, we explore the relation between the focused state and EEG activities. As such, our investigation is oriented towards assessing the viability of identifying a focused state through binocular focusing point disparity. Concurrently, our inquiry delves into the potential existence of an implicit mapping between EEG signal features and focused states and confirms its generalizability.

The primary contributions of this paper can be succinctly outlined as follows:

\begin{itemize}
    \item We propose an annotation method for focused states by computing binocular focusing point disparity with eye movement signals.
    \item The integration of raw eye movement signals with EEG signal data obtained from bio-acquisition devices has been executed to construct a comprehensive dataset with the focused labels and the EEG features.
    \item Several deep learning models are employed to classify the focused states from EEG, yielding impressive performance and affirming the robustness of our approach. Cross-subject experiments are conducted to validate the generalizability of this method.
\end{itemize}

\section{METHOD}
\begin{figure}[!htb]
    \centering
    \includegraphics[width=1\linewidth]{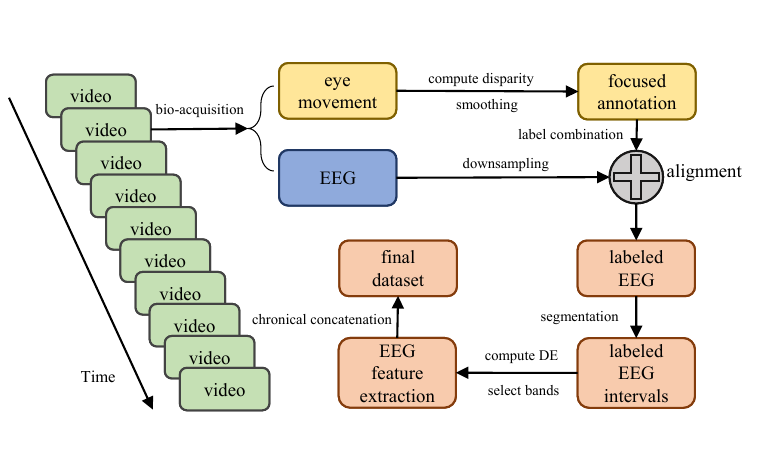}
    \caption{ The pipeline of whole data process.}
    \label{fig:pipeline}
\end{figure}

The comprehensive data processing flowchart for this experiment is illustrated in Fig. \ref{fig:pipeline}, and the pivotal steps are delineated individually below.

\subsection{Experiment Setup}
We used the EEG and the eye movement signal dataset constructed in the experiment by Jiang \textit{et al.}\cite{dataset}. In their experimental procedure, subjects sequentially viewed ten video segments while their eye movement signals and EEG signals were recorded using a bio-acquisition equipment. The EEG signals were obtained from 62 channels, accompanied by corresponding timestamps and behavior event labels for each time segment (such as the start and end times of each video segment, the ID of the viewed video, etc.). The eye movement data recorded the coordinates of the gaze points, gaze duration, and eye movement behaviors for both eyes of the subjects. The data also included timestamps and behavior event labels for each segment (similar to EEG). In the experiments, a total of seventeen participants, aged between 18 and 30, were recruited, comprising 9 males and 8 females. Each participant underwent the experimental procedure three times.

Throughout the experimental phase, an incidence of missing eye movement data emerged, which could be attributed to the unstable sampling of the device. To mitigate this issue, we chose to employ statistical methods to calculate the percentage of missing values within the experimental data for each subject. Consequently, we excluded subjects exhibiting a substantial number of missing values. To maintain data continuity, we applied interpolation to the remaining data to fill in the limited missing values.

\subsection{Annotation of Focused State}
As outlined in Section \ref{sec:Introduction}, our approach involves utilizing eye movement signals to ascertain a focused state. The initial step entails processing raw eye movement signal data obtained from each subject. This involves segmenting the data into ten periods corresponding to the chronological order of watching the ten videos, based on the behavioral event label and timestamp.
Subsequently, determine the focused state of the eye movement signal by assessing the binocular focusing point disparity between the left and right eyes. This determination is made while retaining the \textit{GazeEventType} attribute, which encompasses fixation, saccade, and unclassified events.

The binocular disparity is minimized in a focused state, contrasting with a larger disparity observed in an unfocused state. Accounting for systematic errors and inter-individual differences, we employ a judgment method that combines both the standard deviation and the mean of the binocular disparity in each video. This method can be described as:

\begin{align}
    \textit{State} = \begin{cases}
        \textit{Focused}\ \ &\text{if } disparity \leq mean +std. \\ 
        \textit{Unfocused}\ \ &\text{otherwise}.
    \end{cases}
\end{align}

Upon obtaining the focused label, significant fluctuations are observed. To address this, we opt for correction using the previously retained \textit{GazeEventType} metric. Specifically, consider moments as truly focused only when they are in a fixation situation and judged to be focused (as rapid eye movements couldn't be distinguished from sweeping or unconscious eye movements). Then merge two labels to derive the final focused label, which will serve as classification labels for EEG features.

To ensure smoother intervals of focused and unfocused states, considering fluctuations caused by noise, and to prevent excessively short time intervals from impeding subsequent EEG feature extraction, thereby leading to an increase in missing values in EEG features, we employ a smoothing approach by consolidating brief intervals into adjacent longer ones. The strategy here we apply is to convert excessively short unfocused intervals into focused ones, as subjects are objectively expected to conscientiously engage in emotional experiments.

\subsection{EEG Feature Extraction}
\label{imb}
Similar to the eye movement signals, for each subject, we first extract the EEG signals (62 channels) corresponding to the period during which the 10 videos are viewed, based on the behavioral event labels and associated timestamps. Due to differences in sampling rates between the EEG acquisition device and eye movement signals, we downsample the EEG signals to a uniform frequency for alignment purposes.

Subsequently, we chronologically merge the EEG signals with the corresponding focused labels and segment the EEG data into consecutive focused and unfocused time slices according to the labels. For each focused or unfocused time slice within the same video, we compute the Differential Entropy (DE) features for each channel in that time slot. Additionally, we extract features from five frequency bands (i.e., delta: 1-3 Hz, theta: 4-7 Hz, alpha: 8-13 Hz, beta: 14-30 Hz, and gamma: 31-50 Hz). The differential entropy is given by:
\begin{align}
    \textit{DE} &= log_2(\frac{\sum|\textit{FFT}(data_i)|^2}{end-start+1}). 
\end{align}
Ultimately, we obtain the final 62*5 EEG features for each time slice, and a sample of the final dataset is created by appending the focused label. To maintain temporal ordering, we sequentially merge all time slices in each video from each subject after feature extraction, resulting in a complete dataset for each video (10 in each subject).

It is important to note that, given the experimental requirements dictating subjects to be in a focused state most of the time, the final focused and unfocused samples were unbalanced, with more focused samples than unfocused samples. This imbalance could potentially impact the subsequent classification by machine learning models. To address this, we opt to randomly upsample the unfocused samples in the training set and downsample the positive samples randomly, ensuring a balanced distribution of positive and negative samples. The original dataset is retained for testing to validate the classification performance.

\subsection{Classifier Selection}
\label{data}
To account for inter-individual differences in thinking patterns, we adopt a subject-dependent training approach. For each subject, randomly shuffle the dataset of the ten videos at the video level and divide it into training and test sets in a 7:3 ratio. Subsequently, combine all samples from these sets, obtain distinct training and test sets for each subject, and perform sample balance processing on the training set. To ensure stability, we repeat this process 20 times, averaging the results to assess the performance on each subject of each model. Following this, we calculate the means and standard deviations of the performance across all subjects under the same model to derive the overall average performance of the model. This approach is taken to identify stable patterns and ensure a robust evaluation of the effectiveness of the model.

We utilize six models, namely Support Vector Machine (SVM), Logistic Regression (LR), Multi-layer Perceptron (MLP), Convolutional Neural Network (CNN), Recurrent Neural Network (RNN), and Transformer introduced by Vaswani \textit{et al}\cite{transformer} which applies utilization of the self-attention mechanism, allowing the model to simultaneously focus on all positions in a sequence when processing sequence data. The first three models stretch the 62*5 features of the sample into a line of length 310. The MLP has a total of three fully connected layers; the CNN contains three 2D convolutional layers, a maximum pooling layer, and two fully connected layers; the RNN contains two bi-directional recurrent hidden layers and two fully connected layers; for the Transformer here we set the configuration to include 8 encoder modules, with 2 attention heads in the multi-head attention mechanism, and dimensions of 16 for both the heads and the MLP. We uniformly choose the same Adam optimizer, learning rate, and the number of iterations, and set the batch size to 64. Cross entropy is applied as the loss function. The respective optimal model parameters were chosen based on a large number of experiments.
\section{EXPERIMENT \& RESULT DISCUSSION}

\subsection{Results Analysis}
In order to explore the effect of the five frequency bands on the model performance, we also take out each of the five bands and train them individually, controlling all other hyperparameters to be exactly the same and comparing the performance with all bands. The results are shown in the Table \ref{accs}.
\begin{table*}[!htb]
    \centering
    \caption{Classification accuracy (mean/std)\% of different models.}
    \begin{tabular}{p{0.1\textwidth}|p{0.11\textwidth}|p{0.11\textwidth}|p{0.11\textwidth}|p{0.11\textwidth}|p{0.11\textwidth}|p{0.11\textwidth}}
    \hline
\textbf{Models} & \textbf{delta} & \textbf{theta} & \textbf{alpha} & \textbf{beta} & \textbf{gamma} & \textbf{all}\\
        \hline
SVM & 81.84/9.73 & 78.75/11.26 & 80.13/10.88 & 77.35/13.92 & 80.79/10.40 & 82.29/9.25\\
        \hline
LR & 84.76/7.45 & 84.01/9.42 & 82.17/10.53 & 83.59/9.04 & 84.89/8.40 & 85.05/8.13\\
        \hline
MLP & 84.37/6.95 & 81.14/11.36 & 82.26/7.18 & 83.19/8.02 & 84.61/7.00 & 85.27/6.88\\
        \hline
CNN & 87.14/6.16 & 85.33/7.94 & 81.15/7.26 & 86.81/6.90 & 86.99/6.75 & 87.34/6.25\\
        \hline
RNN & 89.17/6.80 & 86.36/9.21 & 88.88/7.57 & 88.96/6.80 & 89.04/7.31 & 89.23/6.77\\
    \hline
\textbf{Transformer} & \textbf{89.43/4.81} & \textbf{89.18/4.92} & \textbf{88.96/5.22} & \textbf{89.37/4.64} & \textbf{89.50/4.97} & \textbf{90.16/4.62}\\
    \hline
    \end{tabular}
    \label{accs}
\end{table*}
From the results, it can be seen that Transformer has the best performance in terms of accuracy and stability.

The quantitative differences vary due to the different sample sizes of focused versus unfocused samples, as well as inter-individual differences in subjects. For the subject with more focused samples, simple models such as SVM can not be able to effectively learn the features of the two types of samples resulting in low and unstable performance even with balanced sampling of the training set; the MLP model shows large fluctuations due to its simplicity and difficulty in modeling complex coding and classification through fully connected layers and activation functions; whereas the Transformer comprehensively learns the relationship between the channels and the frequency bands and captures more effective information. 

\begin{figure}[!htb]
    \centering
    \subfloat[SVM]
  {
\label{fig:subfig1}\includegraphics[width=0.49\linewidth]{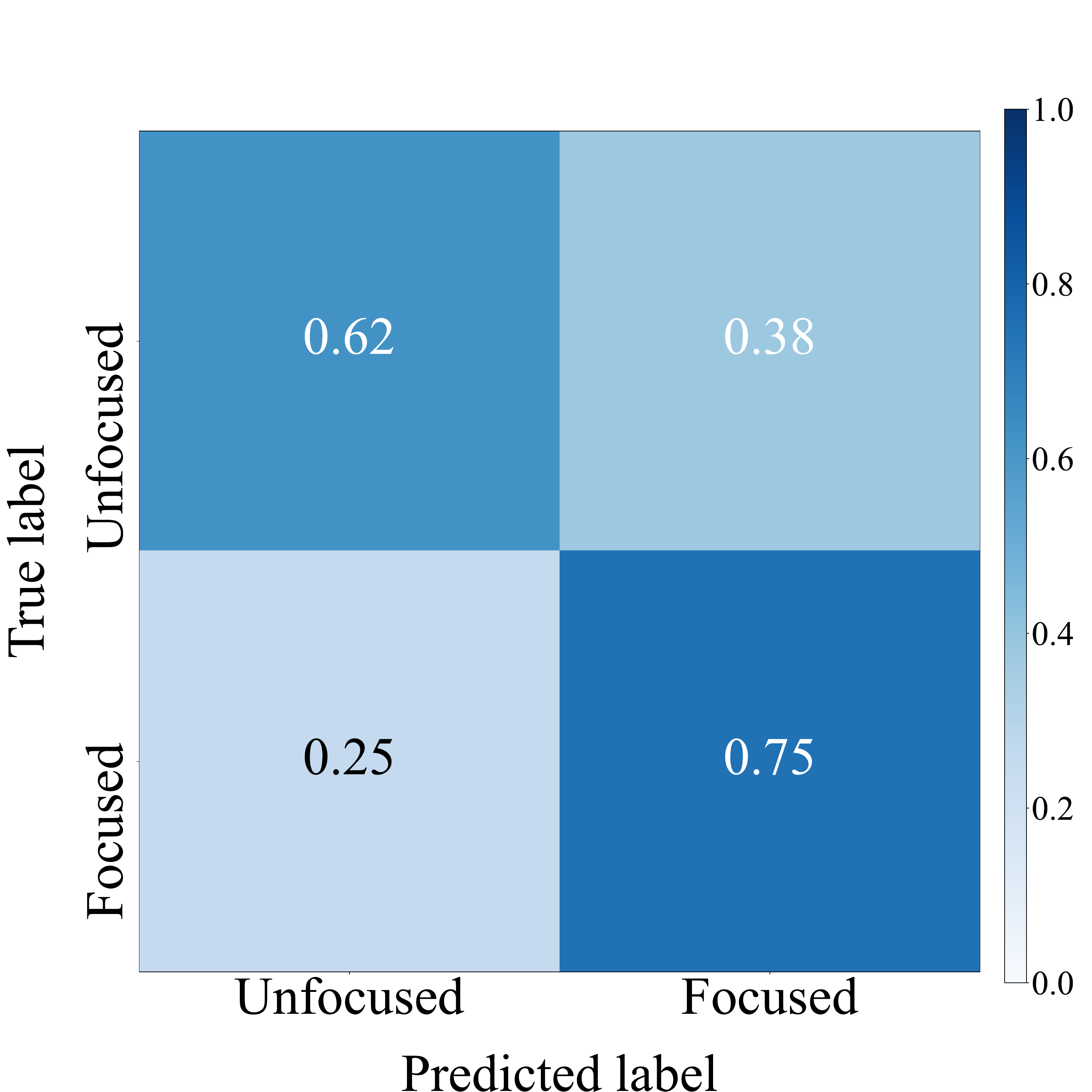}
  }
  \subfloat[Transformer]
  {
\label{fig:subfig2}\includegraphics[width=0.49\linewidth]{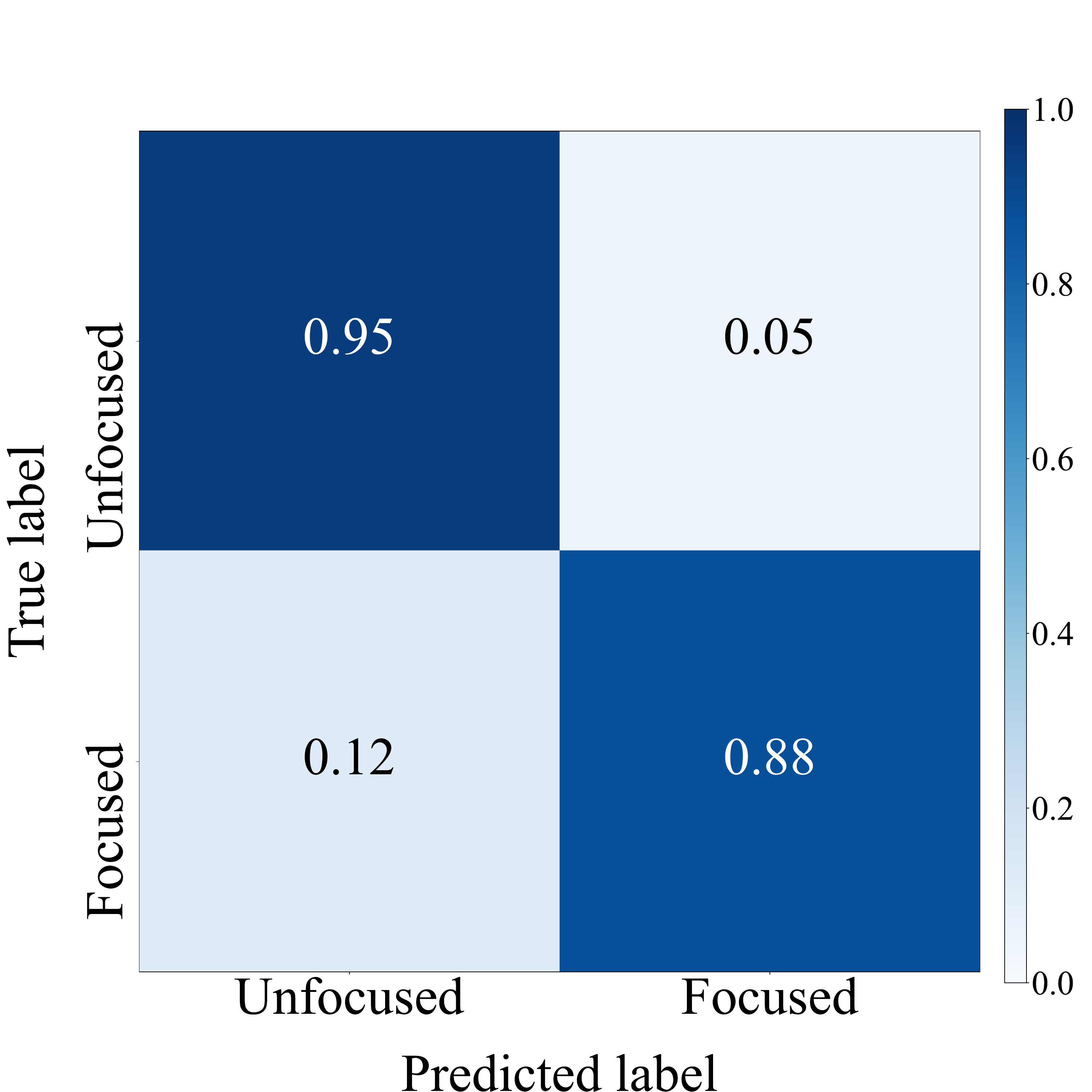}
  }
    \caption{Confusion matrix of SVM and Transformer.}
    \label{fig:transformer}
\end{figure}

To verify the performance of Transformer, we plot the confusion matrix of SVM and Transformer in Fig. \ref{fig:transformer}.
It is not difficult to observe that in Transformer, cases where one class is predicted as another are relatively rare, but compared to this, the frequency of such occurrences is much higher in SVM. This proves that Transformer has the strongest performance in the classification. We can thus consider the annotation method based on eye movement signals as the basis for focused state determination to be feasible and highly reliable. Meanwhile, compared with the performance of each band, there is little difference in the performance across the five single bands, but the delta and gamma bands have higher accuracies than the other bands. None of the single frequency bands are as effective as the whole, indicating that the five bands together carry important EEG features.

\begin{figure}[!htp]
    \centering
    \includegraphics[width=1\linewidth]{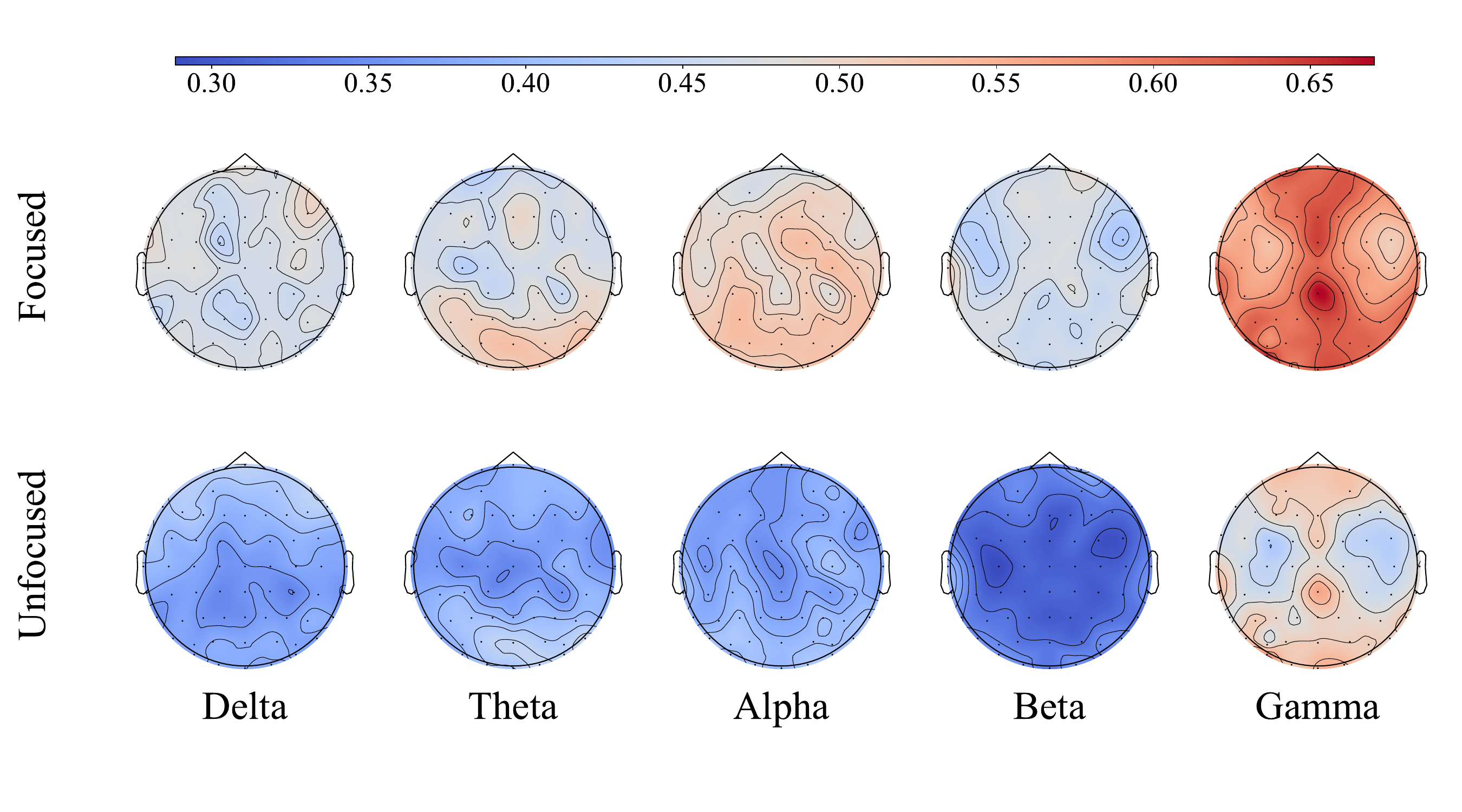}
    \caption{ Topographic maps of the focused and unfocused states in the five frequency bands over all subjects.}
    \label{fig:brain}
\end{figure}

To explore key brain regions under focused versus unfocused states, we investigate the neural topographic maps in order to further understand the characteristics of EEG, as shown in Fig. \ref{fig:brain}.
The topographic map depicts that during a focused state, brain activation energy across all frequency bands significantly surpasses that of an unfocused state. This heightened energy is attributed to increased information acquisition and sustained cognitive activity during focused states, while lower energy levels in all frequency bands during unfocused states suggest a more relaxed or fatigued mental state. Comparison among the five frequency bands reveals that the gamma band exhibits the highest energy, indicative of intense cognitive engagement, followed by theta and alpha bands. Focused states exhibit heightened activations in the parietal region under the gamma band and in the occipital region in the gamma, theta, and alpha bands.
\subsection{Cross-Subject Experiment}
To examine the influence of inter-subject differences on the model performance, we conduct cross-subject experiments using the leave-one-out cross-validation method. In this approach, the same model is trained concurrently using datasets from multiple subjects, and the results are compared with those obtained from the subject-dependent experiments. The Transformer model is utilized for this comparative analysis, and the corresponding Acc, F1-score, and AUC values are presented in Table \ref{cross}.
\begin{table}[!htp]
    \centering
    \caption{Performance (mean/std)\% of the cross-subject and subject-dependent experiments.}
    \begin{tabular}{c|c|c|c}
    \hline
\textbf{Setup} & \textbf{Acc} & \textbf{F1-score} & \textbf{AUC} \\
        \hline
\textbf{cross-subject} & 87.65/5.22 & 90.47/4.98 & 93.18/5.15 \\
    \hline
\textbf{subject-dependent} & 90.16/4.62 & 92.59/4.37 & 94.09/5.06\\
    \hline
    \end{tabular}
    
    \label{cross}
\end{table}

The performance of the cross-subject model exhibits a slight degradation. This decline can be attributed to inter-individual differences, which may introduce variations in feature distributions across subjects. Despite this, the performance remains at a notably high level. This outcome underscores the robustness and generalizability of the focused determination method employed in this study.

\section{CONCLUSION}

This paper presents an annotation method for discerning the focused state through binocular disparity. An accuracy of 90.16\% and F1-score of 92.59\% and AUC of 94.09\% on the Transformer model are obtained for focused state recognition using EEG. Examination of topographic maps exposes heightened brain activations in the gamma band during focused states. Despite performance degradation due to inter-individual differences in the cross-subject experiments, the model sustains a good performance with an accuracy of 87.65$\pm$5.22\%, validating the generalizability of the annotation and the prediction methods.






\begin{thebibliography}{10}

\bibitem{0} C. Lin, C. Zhang, J. Xu, R. Liu, Y. Leng and C. Fu, "Neural Correlation of EEG and Eye Movement in Natural Grasping Intention Estimation," in IEEE Transactions on Neural Systems and Rehabilitation Engineering, vol. 31, pp. 4329-4337, 2023, doi: 10.1109/TNSRE.2023.3327907.

\bibitem{1} M. Phothisonothai, N. Pannurat, K. Orkphol, T. Sooknuan, S. Kampeephat and S. Tantisatirapong, "Cognitive Performance Evaluation in Early Stage of Dementia: A Hybrid EEG/Eye Movement Analysis," 2023 27th International Computer Science and Engineering Conference (ICSEC), Samui Island, Thailand, 2023, pp. 453-456, doi: 10.1109/ICSEC59635.2023.10329676. 

\bibitem{2} W. -L. Zheng, W. Liu, Y. Lu, B. -L. Lu and A. Cichocki, "EmotionMeter: A Multimodal Framework for Recognizing Human Emotions," in IEEE Transactions on Cybernetics, vol. 49, no. 3, pp. 1110-1122, March 2019, doi: 10.1109/TCYB.2018.2797176.

\bibitem{3} T. -H. Li, W. Liu, W. -L. Zheng and B. -L. Lu, "Classification of Five Emotions from EEG and Eye Movement Signals: Discrimination Ability and Stability over Time," 2019 9th International IEEE/EMBS Conference on Neural Engineering (NER), San Francisco, CA, USA, 2019, pp. 607-610, doi: 10.1109/NER.2019.8716943.

\bibitem{4} W. Liu, J. -L. Qiu, W. -L. Zheng and B. -L. Lu, "Comparing Recognition Performance and Robustness of Multimodal Deep Learning Models for Multimodal Emotion Recognition," in IEEE Transactions on Cognitive and Developmental Systems, vol. 14, no. 2, pp. 715-729, June 2022, doi: 10.1109/TCDS.2021.3071170.

\bibitem{5} J. Zhao and Q. Wang, "Eye Movement Attention Based Depression Detection Model," 2022 IEEE 9th International Conference on Data Science and Advanced Analytics (DSAA), Shenzhen, China, 2022, pp. 1-2, doi: 10.1109/DSAA54385.2022.10032433.

\bibitem{6} S. Cheng, J. Wang, L. Zhang and Q. Wei, "Motion Imagery-BCI Based on EEG and Eye Movement Data Fusion," in IEEE Transactions on Neural Systems and Rehabilitation Engineering, vol. 28, no. 12, pp. 2783-2793, Dec. 2020, doi: 10.1109/TNSRE.2020.3048422.

\bibitem{7} A. J. Casson and E. V. Trimble, "Enabling Free Movement EEG Tasks by Eye Fixation and Gyroscope Motion Correction: EEG Effects of Color Priming in Dress Shopping," in IEEE Access, vol. 6, pp. 62975-62987, 2018, doi: 10.1109/ACCESS.2018.2877158.

\bibitem{dataset} W. -B. Jiang, L. -M. Zhao, P. Guo and B. -L. Lu, "Discriminating Surprise and Anger from EEG and Eye Movements with a Graph Network," 2021 IEEE International Conference on Bioinformatics and Biomedicine (BIBM), Houston, TX, USA, 2021, pp. 1353-1357, doi: 10.1109/BIBM52615.2021.9669637.

\bibitem{transformer} A. Vaswani, N. Shazeer, N. Parmar, J. Uszkoreit, L. Jones, A. N. Gomez, L. Kaiser, and I. Polosukhin, "Attention is All You Need", https://arxiv.org/abs/1706.03762

\end{thebibliography}
\end{document}